\documentstyle[preprint,aps]{revtex}
\begin{document}
\title{Sound in relativistic superfluid with vorticity}
\author{G.V. Vlasov\thanks{
E-mail: vs@itp.ac.ru}}
\address{{\it Moscow Aviation Institute and }\\
{\it Landau Institute for Theoretical Physics }\\
2 Kosygin Street, Moscow 117334, Russia }
\maketitle

\begin{abstract}
We develope a theory of sound in a relativistic superfluid with quantum
vortices. The vortices are presented by vortex fluid. For a particular
separable model we find new modes of which a non-relativistic superfluid is
deprived.
\end{abstract}

\draft

\sloppy

\section{Introduction}

{\bf 1}. The basis of relativistic superfluid mechanics was emphasized by
Israel \cite{Israel81} and Dixon \cite{Dixon82} and their achievements were
developed by several authors \cite{KL82,CK92a,CK92b,Q94,CL94,DB95,CL95a}.
Since the relativistic superfluid matter of neutron stars, being the main
object of applications of this theory, contains quantum vortices, an ample
discussion of a relativistic superfluid vortex prompted in diverse forms 
\cite{Davies90,BY91,BY92,BY94,BY95,CL95b} and taken into account by Carter
and Langlois \cite{CL95b} in a new method designed for general hydrodynamic
description of a relativistic superfluid with quantum vortices \cite
{Carter94,CL95c,CLP97,CL98}. For the previous discussion did not explicitly
include vorticity whose presence changes the equations of motion \cite{CL95b}%
. The latter model shall be applied in the present study to investigation of
the sound propagation through a superfluid containing quantum vortices. It
should be noted that the sound \cite{Vilchinskii94,CL95a,BY95} and shock
wave propagation \cite{Vlasov98} in a two-constituent relativistic
superfluid have been already discussed (as well as the recent analysis of
the Cauchy problem \cite{Linet99}); the sound in a ''vorticized''
relativistic superfluid, however close it seems to the usual first-second
sound, bears principal difference from the former, because vorticity $w_{\nu
\rho }$ is interpreted as a tensor dynamical variable on which the
Lagrangian function depends \cite{Carter94}. So, in the model which we shall
exploit below \cite{Carter94,CL95c} the role of the second constituent is
played by the vorticity considered as a vortex ''liquid'' rather than a
vortex lattice where the so-called Tkachenko-Dyson waves \cite{Sonin87} can
be observed (they are, evidently, inconceivable in the present model). Thus,
we consider the waves in a relativistic superfluid in the ''continuous'' or
hydrodynamic limit.

We use a natural system of units $\hbar =c=1$ and a metric~$g_{\nu \rho }=$%
diag$(-1,\,1,\,1,\,1)$.

\section{Relativistic superfluid dynamics with quantum vortices}

The equations of a relativistic superfluid with quantum vortices \cite{CL95c}
include the equation of motion 
\begin{equation}
j^\nu \,w_{\nu \sigma }=0  \label{motion}
\end{equation}
and the conservation law 
\begin{equation}
\nabla _\nu \,n^\nu =0  \label{conserv-n}
\end{equation}
of the particle number current $n^\nu $. The total current 
\begin{equation}
j^\nu =\,n^\nu +\nabla _\sigma \/\lambda ^{\nu \sigma }  \label{j}
\end{equation}
is also conserved: 
\[
\nabla _\nu \,j^\nu =0 
\]
The vorticity 2-form 
\begin{equation}
w_{\nu \rho }=2\bigtriangledown _{[\nu }\,\mu _{\rho ]}  \label{form-2}
\end{equation}
satisfies the closure condition (square brackets imply antisymmetrization of
indices) 
\begin{equation}
\nabla _{[\alpha }\,w_{\nu \rho ]}=0  \label{closure}
\end{equation}
but it is not trivially zero, for the vortices are presented.

Since the current $n^\nu $ and the vorticity $w_{\nu \rho }$ are considered
as new dynamic variables, a variation of the Lagrangian $L$ is given by
formula \cite{CL95c} 
\begin{equation}
d\/L=\mu _\nu \,d\/n^\nu -\frac 12\/\lambda ^{\nu \rho }\,d\/w_{\nu \rho }
\label{Lagrangian}
\end{equation}
Instead of the proper Lagrangian $L$ we can use the pressure function 
\begin{equation}
\Psi =L-\mu _\nu \/n^\nu  \label{pressure}
\end{equation}
whose variation 
\begin{equation}
d\/\Psi =-n^\nu \,d\/\mu _\nu -\frac 12\/\lambda ^{\nu \rho }\,d\/w_{\nu
\rho }  \label{press}
\end{equation}
is expressed in terms of new primary variables $\mu _\nu $ and $w_{\nu \rho
} $. In general the pressure depends on three invariants 
\begin{equation}
\mu ^2=-\mu ^\nu \/\mu _\nu \qquad W^2=\frac 12w^{\nu \rho }\/w_{\nu \rho
}\qquad h^2=h^\nu h_\nu \qquad h^\nu =\,\frac 12\varepsilon ^{\nu \alpha
\beta \gamma }\mu _\alpha w_{\beta \gamma }  \label{inv}
\end{equation}
%%%%%%%%%%%%%%%%%%%%%%%%%%%%%%
%We use capital 'W' for the vorticity amplitude  
%(instead of small 'w') because we can mix w^2 (square amplitude) 
%and w^{12} (a component of matrix w^{\nu \rho}).
%%%%%%%%%%%%%%%%%%%%%%%%%%%%%%
with helicity vector $h^\nu $, whose conservation 
\begin{equation}
\nabla _\nu \,h^\nu =0  \label{helicity}
\end{equation}
is a consequence of Eq. (\ref{form-2}) and (\ref{closure}). The secondary
variables $n^\nu $ and $\lambda ^{\nu \rho }$ are expressed through primary
variables by relation 
\begin{equation}
\left( 
\begin{array}{c}
n^\nu \\ 
\lambda ^{\nu \rho }
\end{array}
\right) =\left( 
\begin{array}{cc}
\bar F^{\nu \alpha } & \bar Q^{\nu \beta \gamma } \\ 
\bar R^{\alpha \nu \rho } & \bar G^{\nu \beta \rho \gamma }
\end{array}
\right) \left( 
\begin{array}{c}
\mu _\alpha \\ 
w_{\beta \gamma }
\end{array}
\right)  \label{pr-sc}
\end{equation}
with 
\begin{equation}
\bar F^{\nu \alpha }=2\frac{\partial \Psi }{\partial \mu ^2}g^{\nu \alpha
}\qquad \qquad \bar Q^{\nu \beta \gamma }=-2\frac{\partial \Psi }{\partial
h^2}h_\kappa \varepsilon ^{\kappa \nu \beta \gamma }  \label{ncoeff}
\end{equation}
\begin{equation}
\bar R^{\alpha \nu \rho }=-2\frac{\partial \Psi }{\partial h^2}h_\kappa
\varepsilon ^{\kappa \alpha \nu \rho }\qquad \qquad \bar G^{\nu \beta \rho
\gamma }=-2\frac{\partial \Psi }{\partial W^2}g^{\nu \beta }g^{\rho \gamma }
\label{lcoeff}
\end{equation}
It should be emphasized that the vorticity 2-form $w_{\nu \rho }$ does not
reflect the fine structure of the vortex cell, since the system of vortices
is considered as a vortex ''liquid''. All variables are averaged over the
vortex cell and vary over the macroscopic range. In the non-relativistic
limit only space components of the vorticity form $w_{\nu \rho }$ survive,
while $w_{\nu 0}\equiv 0$.

\section{Infinitesimal discontinuities}

Let $\Gamma ^\nu $ be a space-like normal to the front of infinitesimal
discontinuity. We shall apply the so-called Hadamard technique \cite
{Carter89} for investigation propagation of small-amplitude perturbations.
Recently Carter and Langlois \cite{CL95a} succeeded in calculating the first
and second sound speed in a relativistic two-constituent superfluid. We use
the same method to find the sound speed in a superfluid containing quantum
vortices. The infinitesimal deviation $\hat A$ of an arbitrary quantity $A$
is proportional to the deviation of its gradient 
\begin{equation}
\left[ \nabla _\nu \,A\right] =\hat A\Gamma _\nu \/  \label{hadamard}
\end{equation}
Applying formula (\ref{hadamard}) to the conservation laws (\ref{helicity})
and (\ref{conserv-n}) we get, respectively 
\begin{equation}
\Gamma _\nu \,\hat h^\nu =0  \label{hel}
\end{equation}
\begin{equation}
\Gamma _\nu \,\hat n^\nu =0  \label{cons}
\end{equation}
The closure condition (\ref{closure}) gives equation 
\begin{equation}
\Gamma _{[\alpha }\hat w_{\nu \rho ]}=0  \label{clos}
\end{equation}
while the equation of motion (\ref{motion}) and (\ref{form-2}) in the linear
approximation yield 
\begin{equation}
\/n^\nu \Gamma _{[\nu }\hat \mu _{\rho ]}+\Gamma _\sigma \/\hat \lambda
^{\nu \sigma }w_{\nu \rho }=0\,\,  \label{mot}
\end{equation}

A relationship between the deviations of the primary and secondary variables
is established by the formula 
\begin{equation}
\left( 
\begin{array}{c}
\hat n^\nu \\ 
\hat \lambda ^{\nu \rho }
\end{array}
\right) =\left( 
\begin{array}{cc}
F^{\nu \sigma } & Q^{\nu \eta \vartheta } \\ 
R^{\nu \rho \sigma } & G^{\nu \rho \eta \vartheta }
\end{array}
\right) \left( 
\begin{array}{c}
\hat \mu _\sigma \\ 
\hat w_{\eta \vartheta }
\end{array}
\right)  \label{matrix1-2}
\end{equation}
obtained by differentiation of equations (\ref{pr-sc})-(\ref{lcoeff}) where 
\begin{eqnarray}  \label{coefF}
F^{\nu \sigma } &=&-2\frac{\partial \Psi }{\partial \mu ^2}g^{\nu \sigma }-4%
\frac{\partial ^2\Psi }{\partial \left( \mu ^2\right) ^2}\mu ^\nu \mu
^\sigma +4\frac{\partial ^2\Psi }{\partial \mu ^2\partial h^2}h_\kappa
w_{\beta \gamma }\left( \varepsilon ^{\kappa \nu \beta \gamma }\mu ^\sigma
+\varepsilon ^{\kappa \sigma \beta \gamma }\mu ^\nu \right)  \nonumber \\
&&\ \ -4\frac{\partial ^2\Psi }{\partial \left( h^2\right) ^2}h_\kappa
\varepsilon ^{\kappa \sigma \lambda \xi }w_{\lambda \xi }h_\alpha
\varepsilon ^{\alpha \nu \beta \gamma }w_{\beta \gamma }-2\frac{\partial
\Psi }{\partial h^2}\varepsilon ^{\kappa \nu \beta \gamma }\varepsilon
_\kappa ^{\sigma \lambda \xi }w_{\beta \gamma }w_{\lambda \xi }
\end{eqnarray}
\begin{eqnarray}  \label{coefQ}
Q^{\nu \eta \vartheta } &=&4\frac{\partial ^2\Psi }{\partial \mu ^2\partial
W^2}\mu ^\nu w^{\eta \vartheta }+4\frac{\partial ^2\Psi }{\partial \mu
^2\partial h^2}h_\kappa \varepsilon ^{\kappa \alpha \eta \vartheta }\mu
_\alpha \mu ^\nu -4\frac{\partial ^2\Psi }{\partial h^2\partial W^2}h_\kappa
\varepsilon ^{\kappa \nu \alpha \beta }w_{\alpha \beta }w^{\eta \vartheta } 
\nonumber \\
&&\ \ -\varepsilon ^{\kappa \alpha \eta \vartheta }\mu _\alpha w_{\beta
\gamma }\left( 4\frac{\partial ^2\Psi }{\partial \left( h^2\right) ^4}%
\varepsilon ^{\lambda \nu \beta \gamma }h_\kappa h_\lambda +2\frac{\partial
\Psi }{\partial h^2}\varepsilon _\kappa ^{\nu \beta \gamma }\right) -2\frac{%
\partial \Psi }{\partial h^2}h_\kappa \varepsilon ^{\kappa \nu \eta
\vartheta }
\end{eqnarray}
\begin{equation}
R^{\nu \rho \sigma }=Q^{\sigma \nu \rho }  \label{coefR}
\end{equation}
%%%%%%%%%%%%%%%%%%%%%%%%%%%%%%
%This is obtained by direct calculation and reads as
%$R^{\nu \rho \sigma }=
%=4\frac{\partial ^2\Psi }{\partial W^2\partial \mu ^2}w^{\nu \rho }\mu ^\sigma +
%4\frac{\partial ^2\Psi }{\partial h^2\partial \mu ^2}h_\kappa \varepsilon ^{\kappa \alpha \nu \rho }\mu _\alpha \mu ^\sigma -
%4\frac{\partial ^2\Psi }{\partial W^2\partial h^2}h_\kappa \varepsilon ^{\kappa \sigma \alpha \beta }w_{\alpha \beta }w^{\nu \rho }-
%\varepsilon ^{\kappa \alpha \nu \rho }\mu _\alpha w_{\beta \gamma }\left( 4\frac{\partial ^2\Psi }{\partial \left( h^2\right) ^2}\varepsilon ^{\lambda \sigma \beta \gamma }h_\kappa h_\lambda -2\frac{\partial \Psi }{\partial h^2}\varepsilon _\kappa ^{\sigma \beta \gamma }\right) -
%$2\frac{\partial \Psi }{\partial h^2}h_\kappa \varepsilon ^{\kappa \sigma \nu \rho }$
%%%%%%%%%%%%%%%%%%%%%%%%%%%%%%
\begin{eqnarray}  \label{coefG}
G^{\nu \rho \eta \vartheta } &=&-2\frac{\partial \Psi }{\partial W^2}g^{\nu
\eta }g^{\rho \vartheta }-4\frac{\partial ^2\Psi }{\partial \left(
W^2\right) ^2}w^{\nu \rho }w^{\eta \vartheta }-4\frac{\partial ^2\Psi }{%
\partial W^2\partial h^2}h_\kappa \mu _\alpha \left( \varepsilon ^{\kappa
\alpha \eta \vartheta }w^{\nu \rho }+\varepsilon ^{\kappa \alpha \nu \rho
}w^{\eta \vartheta }\right)  \nonumber \\
&&-\varepsilon ^{\kappa \alpha \nu \rho }\mu _\alpha \mu _\beta \left( 4%
\frac{\partial ^2\Psi }{\partial \left( h^2\right) ^2}h_\kappa h_\lambda
\varepsilon ^{\lambda \beta \eta \vartheta }+2\frac{\partial \Psi }{\partial
h^2}\varepsilon _\kappa ^{\beta \eta \vartheta }\right)
\end{eqnarray}

Substituting (\ref{matrix1-2})-(\ref{coefG}) in (\ref{hel})-(\ref{mot}) we
write ten equations 
\begin{equation}  \label{hel2}
\varepsilon ^{\nu \alpha \beta \gamma }\,\Gamma _\nu \left( \hat \mu _\alpha
w_{\beta \gamma }+\mu _\alpha \hat w_{\beta \gamma }\right) =0
\end{equation}
\begin{equation}  \label{clos2}
\Gamma _{[\alpha }\hat w_{\nu \rho ]}=0
\end{equation}
\begin{equation}  \label{cons2}
\Gamma _\nu \left( F^{\nu \sigma }\hat \mu _\sigma +Q^{\nu \eta \vartheta }%
\hat w_{\eta \vartheta }\right) =0
\end{equation}
\begin{equation}  \label{mot2}
2\/n^\nu \Gamma _{[\nu }\hat \mu _{\sigma ]}+\Gamma _\rho \/\left( R^{\nu
\rho \alpha }\hat \mu _\alpha +G^{\nu \rho \eta \vartheta }\hat w_{\eta
\vartheta }\right) w_{\nu \sigma }=0\,\,
\end{equation}
with ten independent unknowns, namely $\hat \mu _0$, $\hat \mu _1$, $\hat \mu
_2$, $\hat \mu _3$, $\hat w_{01}$, $\hat w_{02}$, $\hat w_{03}$, $\hat w%
_{12} $, $\hat w_{13}$, $\hat w_{23}$. A linear system (\ref{hel2})-(\ref
{mot2}) is consistent if and only if its determinant vanishes, that yielding
the characteristic equation for speed $u$ of the wave.

\section{Dilatonic model}

In order to obtain an explicit analytic solution we shall confine ourselves
with a particular dilatonic model \cite{CL95c} whose Lagrangian $L$ and,
hence, the pressure function $\Psi $ do not depend on the helicity $h$.
Although this model is devoid of complete strictness, for weak vorticity is
provided ($GW^2\ll \mu n$), it corresponds to the conditions inside the
neutron stars and, hence, describes satisfactory the real superfluid matter.
Thus, formulae (\ref{coefF})-(\ref{coefG}) reduces to 
\begin{equation}
F^{\nu \sigma }=-2\frac{\partial \Psi }{\partial \mu ^2}g^{\nu \sigma }-4%
\frac{\partial ^2\Psi }{\partial \left( \mu ^2\right) ^2}\mu ^\nu \mu ^\sigma
\label{FF}
\end{equation}
\begin{equation}
Q^{\nu \eta \vartheta }=4\frac{\partial ^2\Psi }{\partial \mu ^2\partial W^2}%
\mu ^\nu w^{\eta \vartheta }  \label{QQ}
\end{equation}
\begin{equation}
G^{\nu \rho \eta \vartheta }=-2\frac{\partial \Psi }{\partial W^2}g^{\nu
\eta }g^{\rho \vartheta }-4\frac{\partial ^2\Psi }{\partial \left(
W^2\right) ^2}w^{\nu \rho }w^{\eta \vartheta }  \label{GG}
\end{equation}
while (\ref{coefR}) remains the same.

For a dilatonic model it is convenient to practice with a dilaton $\Phi ^2$
instead of the set (\ref{inv}). In the weak vorticity limit \cite{CL95c} we
have 
\begin{equation}
\Psi \left[ \Phi \left( \mu ,W\right) \right] =\Phi ^2\frac{dV\left[ \Phi
\right] }{d\Phi ^2}-V\left[ \Phi \right]  \label{Psi}
\end{equation}
Since 
\begin{equation}
n^2=2\Phi ^4\frac{dV}{d\Phi ^2}+2KW\Phi ^4\qquad \Phi ^{-2}=\frac \mu n
\label{n}
\end{equation}
thereby, 
\begin{equation}
\mu ^2=2\frac{dV}{d\Phi ^2}+2KW  \label{mu}
\end{equation}
and 
\begin{equation}
d\Phi ^2=\left[ d\mu ^2-\frac KW\,dW^2\right] /2V^{\prime \prime }
\label{Phi}
\end{equation}
where 
\begin{equation}
V^{\prime }=\frac{d\,V}{d\,\Phi ^2}  \label{dV}
\end{equation}
and quantity $K$ (whose logarithmic dependance on vorticity is neglected),
being proportional to the circulation quantum round an individual vortex,
satisfies the inequalities 
\begin{equation}
KW\ll V^{\prime }\qquad \qquad \Phi ^2KW\ll V^{\prime \prime }  \label{ineq}
\end{equation}
while the sound speed is given by the formula 
\begin{equation}
c_s^2=\frac n\mu \frac{d\mu }{dn}=\frac{\Phi ^2V^{\prime \prime }}{\Phi
^2V^{\prime \prime }+\mu ^2}  \label{sound}
\end{equation}
In the light of (\ref{Psi})-(\ref{dV}) we obtain the derivatives 
\begin{equation}
\frac{\partial \Psi }{\partial \mu ^2}=\frac{d\Psi }{d\Phi ^2}/\left( \frac{%
\partial \mu ^2}{\partial \Phi ^2}\right) =\frac 12\Phi ^2\left[ \mu
,W\right]  \label{dmu}
\end{equation}
\begin{equation}
\frac{\partial \Psi }{\partial W^2}=\frac{d\Psi }{d\Phi ^2}\left( \frac{%
\partial \Phi ^2}{\partial W^2}\right) =-\frac 12\Phi ^2\left[ \mu ,W\right] 
\frac KW  \label{dw}
\end{equation}
and 
\begin{equation}
\frac{\partial \Psi }{\partial \left( \mu ^2\right) ^2}=\frac 12\frac{%
\partial \Phi ^2}{\partial \mu ^2}=\frac 1{4V^{\prime \prime }}
\label{dmumu}
\end{equation}
\begin{equation}
\frac{\partial \Psi }{\partial \mu ^2\partial W^2}=-\frac K{4WV^{\prime
\prime }}  \label{dmuw}
\end{equation}
\begin{equation}
\frac{\partial \Psi }{\partial \left( W^2\right) ^2}=\frac 14\frac K{W^3}%
\left( \Phi ^2+\frac K{WV^{\prime \prime }}\right)  \label{dww}
\end{equation}

\section{Explicit solution}

Let the vortices be aligned along the axis $z$ and there is no dependence on 
$z$. The only non-zero components will be $w_{01}$, $w_{02}$, $w_{12}$. Also
only $\mu _0$ and $n^0$ are not equal to zero in the local reference frame
commoving the fluid (or one may merely consider the fluid at rest). We can
choose the normal vector $\Gamma ^\nu $ as 
\begin{equation}
\Gamma ^\nu =\left( u,\/\cos \chi ,\/0,\/\sin \chi \right) =\left(
u,\/\Sigma ,\/0,\/\Omega \right)  \label{gamma}
\end{equation}
where $\chi $ is the angle between the axis $z$ and the direction of the
wave propagation. The coefficients (\ref{FF})-(\ref{GG}) and (\ref{coefR}),
then, take the form 
\begin{equation}
F^{\nu \sigma }=-\Phi ^2\,g^{\nu \sigma }-\frac{\mu ^2\,g^{\nu 0}\,g^{\sigma
0}}{V^{\prime \prime }}  \label{F}
\end{equation}
\begin{equation}
Q^{\nu \eta \vartheta }=\Xi \,g^{\nu 0}\,w^{\eta \vartheta }  \label{Q}
\end{equation}
\begin{equation}
G^{\nu \rho \eta \vartheta }\simeq G\/\left( g^{\nu \eta }g^{\rho \vartheta
}-\frac{w^{\nu \rho }w^{\eta \vartheta }}{W^2}\right)  \label{G}
\end{equation}
where $\Xi =\mu \,K/(W\,V^{\prime \prime })$ and $G=\Phi ^2\,K/W$. Note that
in Eq.~(\ref{G}) we do not add the negligible term $K\,W/V^{\prime \prime }$
incorporating Eq.~(\ref{dww}). Substituting formulae (\ref{F})-(\ref{G}) in
equations (\ref{hel2})-(\ref{mot2}) and taking into account (\ref{inv}),
namely 
\begin{equation}
W^2=w^{10}w_{10}+w^{20}w_{20}+w^{12}w_{12}  \label{w2}
\end{equation}
we get 
\begin{equation}
-\hat \mu _0\Omega w_{12}-\hat \mu _1\Omega w_{02}+\hat \mu _2\Omega w_{01}-%
\hat \mu _3U+\hat w_{23}\Sigma \mu _0+\hat w_{12}\Omega \mu _0=0
\label{hel11}
\end{equation}
\begin{equation}
-u\hat w_{12}+\Sigma \hat w_{20}=0  \label{clos22}
\end{equation}
\begin{equation}
-u\hat w_{13}+\Sigma \hat w_{30}+\Omega \hat w_{01}=0  \label{clos3}
\end{equation}
\begin{equation}
-u\hat w_{23}+\Omega \hat w_{02}=0  \label{clos4}
\end{equation}
\begin{equation}
\Sigma \hat w_{23}+\Omega \hat w_{12}=0  \label{clos5}
\end{equation}
\begin{equation}
-u\hat \mu _0\,F^{00}+\Sigma \hat \mu _1F^{11}+\Omega \hat \mu
_3F^{33}-u\left( Q^{001}\hat w_{01}+Q^{002}\hat w_{02}+Q^{012}\hat w%
_{12}\right) =0  \label{cons6}
\end{equation}
\begin{equation}
\hat \mu _0R_0+\hat w_{\eta \vartheta }\,\alpha _0^{\eta \vartheta }=0
\label{zero7}
\end{equation}
\begin{equation}
\hat \mu _0(-2n^0\Sigma +R_1)-\hat \mu _12n^0u+\hat w_{\eta \vartheta
}\,\alpha _1^{\eta \vartheta }=0  \label{first8}
\end{equation}
\begin{equation}
\hat \mu _0\,R_2-2un^0\hat \mu _2\,+\hat w_{\eta \vartheta }\,\alpha
_2^{\eta \vartheta }=0  \label{second9}
\end{equation}
\begin{equation}
\hat \mu _0\Omega \/+\hat \mu _3u=0  \label{third10}
\end{equation}
where 
\begin{equation}
\alpha _j^{\eta \vartheta }=G\,a_j^{\eta \vartheta }  \label{aa}
\end{equation}
\begin{eqnarray}  \label{rrrr}
R_0 &=&\Sigma w_{02}R^{120}-uw_{0k}R^{0k0}\qquad R_2=w_{02}\Sigma
R^{010}+w_{12}uR^{010}\qquad  \nonumber \\
R_1 &=&\Sigma w_{01}R^{010}+w_{12}(\Sigma R^{120}-uR^{020})
\end{eqnarray}
\begin{eqnarray}  \label{a0}
a_0^{01} &=&-B_0\,U\qquad a_0^{02}=-B_0\,U\qquad a_0^{03}=0\qquad  \nonumber
\\
a_0^{12} &=&(1-B_0)\,U\qquad a_0^{13}=-w_{01}\Omega \qquad
a_0^{23}=-w_{02}\Omega
\end{eqnarray}
\begin{eqnarray}  \label{a1}
a_1^{01} &=&B_1\,U\qquad a_1^{02}=(1+B_1)\,U\qquad a_1^{03}=\Omega
w_{01}\qquad  \nonumber \\
a_1^{12} &=&B_1\,U\qquad a_1^{13}=0\qquad a_1^{23}=-\Omega w_{12}
\end{eqnarray}
\begin{eqnarray}  \label{a2}
a_2^{01} &=&(1-B_2)\,U\qquad a_2^{02}=-B_2\,U\qquad a_2^{03}=\Omega
w_{02}\qquad  \nonumber \\
a_2^{12} &=&-B_2\,U\qquad a_2^{13}=\Omega w_{12}\qquad a_2^{23}=0
\end{eqnarray}
while 
\begin{equation}
B_0=\frac{w^{01}\,w^{12}}{W^2}\qquad B_1=\frac{w^{01}\,w^{02}}{W^2}\qquad
B_2=\frac{w^{01}\,w^{01}}{W^2}  \label{b}
\end{equation}
\begin{equation}
U=u\/w_{12}+\Sigma \/w_{02}  \label{m}
\end{equation}
It should be noted that $R_1$ in (\ref{first8}), (\ref{rrrr}) and $Q^{012}$
in (\ref{cons6}) survive in the non-relativistic limit.

\section{Waves}

\subsection{Horizontal waves}

We call the wave propagating along the axis $x$ (i.e. at the right angle to
the axis of vortices $z$) ''horizontal'' since for this case $\Omega =0$, $%
\Sigma =1$. The matrix of the linear system (\ref{hel11}-\ref{third10}),
then, has the form 
\begin{equation}
\left( 
\begin{array}{cccccccccc}
0 & 0 & 0 & -w_{02}-uw_{12} & 0 & 0 & 0 & 0 & 0 & \mu _0 \\ 
0 & 0 & 0 & 0 & 0 & 1 & 0 & u & 0 & 0 \\ 
0 & 0 & 0 & 0 & 0 & 0 & 1 & 0 & u & 0 \\ 
0 & 0 & 0 & 0 & 0 & 0 & 0 & 0 & 0 & u \\ 
0 & 0 & 0 & 0 & 0 & 0 & 0 & 0 & 0 & 1 \\ 
-uF^{00} & F^{11} & 0 & 0 & Q^{01} & Q^{02} & 0 & Q^{12} & 0 & 0 \\ 
R_0 & 0 & 0 & 0 & \alpha _0^{01} & \alpha _0^{02} & 0 & \alpha _0^{12} & 0 & 
0 \\ 
-1+\tilde R_1 & -u & 0 & 0 & \tilde \alpha _1^{01} & \tilde \alpha _1^{02} & 
0 & \tilde \alpha _1^{12} & 0 & 0 \\ 
\tilde R_2 & 0 & -u & 0 & \tilde \alpha _2^{01} & \tilde \alpha _2^{02} & 0
& \tilde \alpha _2^{12} & 0 & 0 \\ 
0 & 0 & 0 & u & 0 & 0 & 0 & 0 & 0 & 0
\end{array}
\right)  \label{matr-h}
\end{equation}
where for every variable $O$ we introduced the notation $\tilde O=O/(2n)$
and $Q^{\nu \rho }=-uQ^{0\nu \rho }$. The solution follows from a condition
of the determinant (\ref{matr-h}) vanish that implies the possibility of
longitudinal waves ($\hat \mu _1\neq 0$ while $\hat \mu _2=\hat \mu _3=0$,
also $\hat w_{03}=\hat w_{13}=\hat w_{23}=0$) with the relevant speed found
from the characteristic equation 
\begin{equation}
(u^2F^{00}-F^{11})\,U^2\det \left( 
\begin{array}{cc}
-B_0\,\qquad & 1-B_0\left( 1-u\right) \\ 
1-B_2\qquad & B_2\,\left( 1-u\right)
\end{array}
\right) =o(K)  \label{hl}
\end{equation}
which, according to Eq.~(\ref{F}), gives the usual sound speed (\ref{sound}%
), for 
\[
F^{00}=-\Phi ^2-\frac{\mu ^2}{V^{\prime \prime }}\qquad F^{11}=-\Phi ^2 
\]
We neglected in (\ref{hl}) the insufficient term $\tilde R_1$ which reflects
the negligible space anisotropy of the first sound; in the relevant equation
for the first sound in vertical direction (see next section) this term is
absent at all. The second solution of (\ref{hl}) is 
\begin{equation}
u_{II}=\frac{w^{20}}{w_{12}}  \label{v}
\end{equation}
In the non-relativistic limit $w_{01}=w_{02}=0$ and the only non-zero
elements of (\ref{a0}-\ref{a2}) will be $a_1^{02}$ and $a_1^{12}$. The
matrix (\ref{matr-h}), then, determines the only ''first sound'' solution,
while (\ref{v}) does not occur, for it pertains to a relativistic superfluid.

\subsection{Vertical waves}

Let the wave propagating along the axis $z$ of the vortex cell, when $\Sigma
=0$ and $\Omega =1$, let be called ''vertical''. Eq. (\ref{clos5}), then,
gives $\hat w_{12}=0$, while the variables $\hat w_{01}$ and $\hat w_{02}$
can be excluded by means of Eqs. (\ref{clos3}) and (\ref{clos4}). Then, we
have to calculate the determinant of matrix

\begin{equation}
\left( 
\begin{array}{ccccccc}
\hat \mu _0 & \hat \mu _1 & \hat \mu _2 & \hat \mu _3 & \hat w_{03} & \hat w%
_{13} & \hat w_{23} \\ 
-w_{12} & -w_{02} & w_{01} & -u\/w_{12} & 0 & 0 & 0 \\ 
-u\/F^{00} & 0 & 0 & F^{33} & 0 & -uQ^{01} & -uQ^{02} \\ 
-1 & 0 & 0 & -u & 0 & 0 & 0 \\ 
R_0 & 0 & 0 & 0 & 0 & \alpha _0^{13}+u\alpha _0^{01} & \alpha
_0^{23}+u\alpha _0^{02} \\ 
\tilde R_1 & -u & 0 & 0 & \tilde Gw_{01} & u\tilde \alpha _1^{01} & \tilde 
\alpha _1^{23}+u\tilde \alpha _1^{02} \\ 
\tilde R_2 & 0 & -u & 0 & \tilde Gw_{02} & \tilde \alpha _2^{13}+u\tilde 
\alpha _2^{01} & u\tilde \alpha _2^{02}
\end{array}
\right)
\end{equation}
where the first row includes the independent variables. The speed of the
longitudinal wave, characterized by $\hat \mu _1=\hat \mu _2=0$, follows
from the equation 
\begin{equation}  \label{vl}
\left( u^2\/F^{00}-F^{33}\right) \det \left( 
\begin{array}{ccc}
0 & b_0^1 & b_0^2 \\ 
w_{01} & b_1^1 & b_1^2 \\ 
w_{02} & b_2^1 & b_2^2
\end{array}
\right) =o(K^4)
\end{equation}
\begin{equation}  \label{b0}
b_0^1=-w_{01}-uB_0U\qquad b_0^2=-w_{02}-uB_0U
\end{equation}
\begin{equation}  \label{b1}
b_1^1=uB_1U\qquad b_1^2=-w_{12}+u\left( 1+B_1\right) U
\end{equation}
\begin{equation}  \label{b2}
b_2^1=w_{12}+u\left( 1-B_2\right) U\qquad b_2^2=-uB_2U
\end{equation}
and coincides approximately with the usual sound speed (\ref{sound}). For
the determinant in (\ref{vl}), as one can check by means of (\ref{b}) and (%
\ref{b0})-(\ref{b2}), does not vanish.

As for the transversal waves ($\hat \mu _3=0$) they are determined by
equation

\begin{equation}  \label{vt}
\det \left( 
\begin{array}{ccccc}
-w_{02} & w_{01} & 0 & 0 & 0 \\ 
0 & 0 & 0 & -uQ^{01} & -uQ^{02} \\ 
0 & 0 & 0 & b_0^1 & b_0^2 \\ 
-u & 0 & w_{01} & \tilde b_1^1 & \tilde b_1^2 \\ 
0 & -u & w_{02} & \tilde b_2^1 & \tilde b_2^2
\end{array}
\right) =0
\end{equation}
which trivially equals zero (due to the matrix degeneracy in (\ref{vt}),
implying impossibility of transversal waves.

\section{Conclusion}

In order to investigate wave propagation in a relativistic superfluid with
quantum vortices we have derived the linear system (\ref{hel2})-(\ref{mot2})
by means of the Hadamard method \cite{Carter89} and have solved the
appropriate characteristic equation. For a dilatonic model in the weak
vorticity limit we found two types of waves propagating at the right angle
to the axis $z$ of vortices. Besides the usual sound (\ref{sound}), an
additional horizontal (in direction orthogonal to $z$) ''second-sound''
branch (\ref{v}) exists, while only the usual sound may propagate in the
direction $z$. It should be noted that no wave propagation with the speed
close to $K\,W/V^{\prime }$ or $\Phi ^2K\,W/V^{\prime \prime }$ were found
(the latter term determines the anisotropy of the first sound in the
horizontal and vertical direction; we do not include this term in Eq. (\ref
{hl}) because it is small and does change the result on qualitative level);
for the second constituent was treated as a vortex ''liquid'', no
transversal waves occur. On the other hand, due to orientation in $z$
direction the condition of wave propagation along $z$ and $x$ (or $y$) is
different. However, only the usual sound may propagate through a
non-relativistic superfluid, and that follows from Eqs. (\ref{hel11}-\ref
{third10}) taken in the non-relativistic limit ($w_{0i}=0$). Because the
matrix of the relevant non-relativistic system 
\begin{equation}
\left( 
\begin{array}{cccccccccc}
\hat \mu _0 & \hat \mu _1 & \hat \mu _2 & \hat \mu _3 & \hat w_{01} & \hat w%
_{02} & \hat w_{03} & \hat w_{12} & \hat w_{13} & \hat w_{23} \\ 
-\Omega w_{12} & 0 & 0 & -u\/w_{12} & 0 & 0 & 0 & \Omega \mu _0 & 0 & \Sigma
\mu _0 \\ 
0 & 0 & 0 & 0 & 0 & -\Sigma & 0 & -u & 0 & 0 \\ 
0 & 0 & 0 & 0 & \Omega & 0 & -\Sigma & 0 & -u & 0 \\ 
0 & 0 & 0 & 0 & 0 & \Omega & 0 & 0 & 0 & -u \\ 
0 & 0 & 0 & 0 & 0 & 0 & 0 & \Omega & 0 & \Sigma \\ 
-u\,F^{00} & \Sigma F^{11} & 0 & \Omega F^{33} & 0 & 0 & 0 & -uQ^{012} & 0 & 
0 \\ 
0 & 0 & 0 & 0 & 0 & 0 & 0 & u\/w_{12} & 0 & 0 \\ 
\Sigma \left( -n^0+\frac 12w_{12}R^{120}\right) & -u & 0 & 0 & 0 & 
\,u\/w_{12} & 0 & 0 & 0 & -\Omega w_{12} \\ 
0 & 0 & -un^0 & 0 & -\,u\/w_{12} & 0 & 0 & 0 & \Omega w_{12} & 0 \\ 
\Omega & 0 & 0 & u & 0 & 0 & 0 & 0 & 0 & 0
\end{array}
\right)  \label{matr-nr}
\end{equation}
yields for the horizontal waves ($\Omega =0$, $\Sigma =1$) the
characteristic equation 
\begin{equation}
u^2F^{00}+F^{11}\left( 1-w_{12}\frac{R^{120}}{2n^0}\right) =0  \label{non1}
\end{equation}
where 
\begin{equation}
R^{120}=\frac{\mu \,K}{W\,V^{\prime \prime }}w^{12}\qquad w_{12}\frac{R^{120}%
}{2n^0}=\frac{\,KW}{2\Phi ^2V^{\prime \prime }}\ll 1  \label{negl}
\end{equation}
in accordance with (\ref{rrrr}), (\ref{Q}), (\ref{QQ}), and merely 
\begin{equation}
u^2F^{00}+F^{33}=0  \label{non2}
\end{equation}
for the vertical waves ($\Omega =1$, $\Sigma =0$). Indeed, Eq. (\ref{non1}),
which we write here in the exact form with the negligible term (\ref{negl}),
follows from (\ref{hl}), while Eq. (\ref{non2}) follows from (\ref{vl}) and
they both determine the usual sound speed (\ref{sound}).

Nevertheless, formula (\ref{v}) requires more discussion. What if (\ref{v})
will be of order of the usual sound (\ref{sound})? Then we have obviously
missed the some information by vanishing $w_{02}$. We do not know the
explicit form of $w_{02}$ and $w_{12}$; although $w_{02}$ vanishes in the
non-relativistic limit, the ratio (\ref{v}) . The analysis of the exterior
region of a vortex line \cite{CL95b} is not sufficient to determine the
components of the vorticity 2-form. These components are generated by the
vortex core and they can be derived from its internal structure; because
outside the vortex core, i.e. outside the support of the vorticity 2-form $%
w_{\nu \rho }$, the irrotationality condition $w_{\nu \rho }=0$ takes place.
So, the complete answer requires explicit derivation of $w_{02}$ and $w_{12}$%
.

Of course, it seems reasonable to discuss in future the shock waves.
However, equations (\ref{hel2})-(\ref{mot2}) have already convey {\it [or: }%
reveal{\it ]} the absence of any other modes except (\ref{sound}) and \ref{v}%
). For instance, the waves through the pure vortex constituent ($\hat \mu
_\nu \equiv 0$) or those similar to the fourth sound \cite{Vlasov98} in
superfluid ($\hat w_{\nu \rho }\equiv 0$) are impossible. This may serve a
hint to search possible shock wave solutions. Moreover, the analysis of a
more complicated model with sufficient vorticity $W$ and the pressure
function $\Psi $ depending on the cross term $h$ may also be performed by
means of Eqs. (\ref{hel2})-(\ref{mot2}).

\end{document}